\documentclass{iopart}
\usepackage{amsfonts}
\usepackage{graphicx}
\usepackage[labelfont=bf]{caption}
\usepackage[hang]{subfigure}
\usepackage[english]{babel}
\usepackage[utf8]{inputenc}
\usepackage[T1]{fontenc}

\usepackage{natbib}
\begin{document}
\title{Analytical approximation of the exterior gravitational field of rotating neutron stars}

\author{C Teichm\"uller$^1$,
M B Fr\"ob$^2$
and F Maucher$^3$}

\address{$^1$ Theoretisch-Physikalisches Institut, University of Jena, Max-Wien-Platz 1, 07743 Jena, Germany}
\address{$^2$ Departament de Física Fonamental, Institut de Ciències del Cosmos (ICC), Universitat de Barcelona, C/Martí i Franquès 1, 08028 Barcelona, Spain}
\address{$^3$ Max Planck Institute for the Physics of Complex Systems, 01187 Dresden, Germany}
\ead{christian.teichmueller@uni-jena.de}

\begin{abstract}
It is known that B\"acklund transformations can be used to generate 
stationary axisymmetric solutions of Einstein's vacuum field equations with any number of constants. 
We will use this class of exact solutions to describe the exterior vacuum region of 
numerically calculated neutron stars.
Therefore we study how an Ernst potential given on the rotation axis and containing an arbitrary number of constants 
can be used to determine the metric everywhere.
Then we review two methods to determine 
those constants from a numerically calculated solution.
Finally, we compare the metric and physical properties 
of our analytic solution with the numerical data and find excellent agreement 
even for a small number of parameters.  
\end{abstract}


\section{Introduction}

Solving Einstein's field equations in the vacuum with the assumption
of axisymmetry and stationarity amounts to little more than solving
Ernst's equation, which is a non-linear partial differential equation.
Ernst's formulation of the stationary axisymmetric problem (\cite{e68a,e68b}, \cite{kn68})
allowed a reformulation of the stationary
axisymmetric Einstein equations in terms of a ``linear system'' (see 
\cite{m78}, \cite{bz78}, \cite{h78}, \cite{n79}), such
that the integrability condition of the latter equals Ernst's equation.

During the past two decades, several papers have been published focussing on
constructing analytical Ernst potentials and applying them to various
settings (e.g. \cite{ms93}, \cite{sc02}, \cite{bs04} and recently \cite{p09}). These include solutions to
generalizations of Ernst's equation, which permit an electromagnetic field
and use given axis data in the form of rational functions with an
arbitrary number of constants on the axis of symmetry based on
Sibgatullin’s method (\cite{s91}).

Here, we will present an alternative approach using Bäcklund transformations 
acting w.l.o.g. on the Minkowski seed solution
(see \cite{n79}, in particular \cite{n80a}) to find concise formulae for
the determination of the involved parameters. This approach will result in
an extremely simple, but in our opinion at the same time elegant recipe,
that does not require any sophisticated mathematical background.

This method will then be used to calculate an analytical solution of the
field equations in vacuum, which approximates the exterior metric for a given model of
a rotating neutron star very accurately.

Two different approaches to compute the axis potential are considered.
The resulting analytical solutions are then compared with rigorous numerical
results using the AKM programme (\cite{akm03}) in the whole vacuum region. 
We find excellent agreement and convergence even
for a small number of parameters.

This paper is organized as follows. In the first part, we review the properties of stationary and axisymmetric
spacetimes and the relation between the metric and Ernst's potential. After that, we introduce the linear problem and
show how to solve it using B{\"a}cklund transformations
with an arbitrary number of constants, referred to as B{\"a}cklund parameters. 
Then we provide concise formulae determining those parameters from the axis potential.
In the next part we illustrate two methods to compute the axis potential, namely by using Geroch-Hansen 
multipole moments and alternatively by fitting
a quotient of polynomials to given axis values. After that we compare our analytical results with
full numerical ones.

\section[Construction of the metric]{Construction of the metric by B\"acklund transformations acting on a seed solution}
\subsection{The Ernst equation}

The exterior vacuum gravitational field of a stationary axisymmetric uniformly rotating
matter configuration can be written in Weyl-Lewis-Papapetrou coordinates\begin{equation}
\mathrm{d}s^{2}=\mathrm{e}^{-2U}\left[\mathrm{e}^{2k}\left(\mathrm{d}\varrho^{2}+\mathrm{d}\zeta^{2}\right)+\varrho^{2}\mathrm{d}\varphi^{2}\right]-\mathrm{e}^{2U}\left(\mathrm{d}t+a\,\mathrm{d}\varphi\right)^{2},\label{eq:metric}\end{equation}
where $U$ denotes the {}``Newtonian'' gravitational potential, $a$ the
``gravitomagnetic'' potential and $k$ the superpotential, cf. e.g. \cite{nm03}.
These three potentials are functions of $\varrho$ and $\zeta$ alone.
The metric (\ref{eq:metric}) admits an Abelian group of motions with
the Killing vectors $\xi^{i}=\delta_{t}^{i}$, $\xi^{i}\xi_{i}\rightarrow-1$ asymptotically as $\varrho^2 +\zeta^2 \rightarrow\infty$
and $\eta^{i}=\delta_{\varphi}^{i}$, $\eta^{i}\eta_{i}>0$ for $\varrho>0$. $\delta_{j}^{i}$
is the Kronecker delta, so that $\xi^{i}$ has only a $t$-component,
and $\eta^{i}$ points in the azimuthal $\varphi$-direction. Note that $U$ and $a$ can be related to scalar products of the Killing vectors by the coordinate independent expressions $\mathrm{e}^{2U}=-\xi^{i}\xi_{i}$
and $a=-\mathrm{e}^{-2U}\eta_{i}\xi^{i}$.

In the vacuum, Einstein's equations for $U$ and $a$ are equal to the so called Ernst equation:\begin{equation}
\Re\left(f\right)\left(f_{,\varrho\varrho}+f_{,\zeta\zeta}+\frac{1}{\varrho}f_{,\varrho}\right)=f_{,\varrho}^{2}+f_{,\zeta}^{2}.\label{eq:ernst}\end{equation}
The metric
coefficients $a,k,U$ and Ernst's potential are related via\[
f\left(\varrho,\zeta\right)=\mathrm{e}^{2U}+\mathrm{i}b,\]
where $b$ was introduced as \begin{equation}\label{eq:b}
a_{,\varrho}=\varrho \mathrm{e}^{-4U}b_{,\zeta},\hspace{0.5cm}a_{,\zeta}=-\varrho \mathrm{e}^{-4U}b_{,\varrho}\end{equation}
and $k$ can be found by\begin{equation}\label{eq:k}
k_{,\varrho}=\varrho\left[U_{,\varrho}^{2}-U_{,\zeta}^{2}+\frac{1}{4}\mathrm{e}^{-4U}\left(b_{,\varrho}^{2}-b_{,\zeta}^{2}\right)\right],\end{equation}\[k_{,\zeta}=2\varrho\left[U_{,\varrho}U_{,\zeta}+\frac{1}{4}\mathrm{e}^{-4U}\left(b_{,\varrho}b_{,\zeta}\right)\right].\]
It follows from (\ref{eq:ernst}), that $a_{,\varrho\zeta}=a_{,\zeta\varrho}$
and $k_{,\varrho\zeta}=k_{,\zeta\varrho}$ are fulfilled, so $a$ and $k$
can be obtained by line integration from $f$. 

\subsection{The linear Problem}
It is well-known (see e.g. \cite{n80b}, \cite{nk83}) that Ernst's equation is the integrability condition
$\Phi_{,z\bar {z} }  =  \Phi_{,\bar {z} z} $ of the linear problem 
\begin{eqnarray}
\Phi_{,z} & = & \left[\left(\begin{array}{cc}
B & 0\\
0 & A\end{array}\right)+\lambda\left(\begin{array}{cc}
0 & B\\
A & 0\end{array}\right)\right]\Phi,\label{eq:linear_system_MN}\\
\Phi_{,\bar{z}} & = & \left[\left(\begin{array}{cc}
\bar{A} & 0\\
0 & \bar{B}\end{array}\right)+\frac{1}{\lambda}\left(\begin{array}{cc}
0 & \bar{A}\\
\bar{B} & 0\end{array}\right)\right]\Phi.\nonumber \end{eqnarray}
Here, we introduced the complex coordinates $z:=\varrho+\mathrm{i}\zeta$, $\bar{z}:=\varrho-\mathrm{i}\zeta$ and the spectral parameter
$\lambda := \sqrt{(K - \mathrm{i}\bar{z})/(K+\mathrm{i}z)}$, where $K$ is an arbitrary complex constant. 
$\Phi(\lambda,\bar{z},z)$
is a $2\times2$    matrix, depending on the coordinates as well as on the spectral
parameter, whereas   $A$ and   $B$  depend only on the coordinates $\varrho$ and $\zeta$. One may
always multiply (\ref{eq:linear_system_MN}) from the right by a so-called gauge matrix which depends only
on $K$.
Equating all coefficients of $\lambda$ in the integrability condition leads to a set of $4$ differential equations. One may introduce a potential $f$ in such a way that 
$B=\bar{f}_{,z}/(f+\bar{f})$ and $A=f_{,z}/(f+\bar{f})$. The remaining two equations
are then the Ernst equation (\ref{eq:ernst}).

\subsection{Solving the linear system}
Solving a linear system, which depends on an additional spectral parameter,
cor\-responds to solving a Riemann-Hilbert problem. 
The idea  is then to discuss $\Phi$ as a holomorphic function of $\lambda$ for fixed $z$ and $\bar{z}$ and 
to calculate $A$ and $B$ afterwards. 
With the special
ansatz that will be discussed in the following, no integrals have to be evaluated. 
Instead, solving the Riemann-Hilbert problem can be reduced to solving a linear system of
equations. This special ansatz is to formulate a set of properties, such that if a
matrix  $\Phi$ has these properties, it automatically fulfils the linear problem.

{\em
A matrix $\Phi$ that has the properties (\ref{eq:P1}-\ref{eq:P4}) is
a solution of the linear system (\ref{eq:linear_system_MN}) (\cite{n96}).
}
\begin{equation}
\Phi_{,z}\Phi^{-1}=Q+\lambda R,\quad\Phi_{,\bar{z}}\Phi^{-1}=S+T/\lambda,\label{eq:P1}\end{equation} where $Q,\,R,\,S$ and $T$ are matrices that do not depend on $\lambda$.
\begin{equation}
\Phi\left(\lambda\right)=\left(\begin{array}{cc}
\psi\left(\lambda\right) & \psi\left(-\lambda\right)\\
\chi\left(\lambda\right) & -\chi\left(-\lambda\right)\end{array}\right)\label{eq:P2}\end{equation}
\begin{equation}
\overline{\psi\left(\bar{\lambda}^{-1}\right)}=\chi\left(\lambda\right)\label{eq:P3}\end{equation}
\begin{equation}
\psi\left(\lambda=-1,z,\bar{z}\right)=\chi\left(\lambda=-1,z,\bar{z}\right)=1\label{eq:P4}\end{equation}
For a detailed discussion see e.g. \cite{m08}.
\cite{nk83} used the ansatz 
\[\Phi = P\Phi_0=\mu\tilde P \Phi_0,\quad\mbox{with } \mu=\left(\frac{K+\rm{i} \textit{z}}{K}\right)^{n}\] to seek new solutions $\Phi$ from old ones $\Phi_0$, where 
\[\tilde P=\sum_{k=0}^{2n} P_k \lambda^k\] is a finite matrix polynomial in $\lambda$. We restrict ourselves w.l.o.g. 
to the Minkowski seed solution
\[\Phi_0 = \left(
\begin{array}{cc}\psi_0 & \psi_0\\ 
\chi_0 & -\chi_0
\end{array}\right) = \left(\begin{array}{cc}1 & 1\\ 
1 & -1
\end{array}\right).\]
The fundamental lemma of algebra gives 
\[
\det \tilde P(\lambda) = \det \sum_{i=0}^{2n} P_i \lambda^i = \beta(z,\bar{z}) \prod_{k=1}^{4n} (\lambda - \lambda_k),
\]
where the zeros $\lambda_k$ are functions of $z$ and $\bar{z}$:
\[
 \lambda_k = \sqrt{\frac{K_k-\mathrm{i}\bar{z}}{K_k+\mathrm{i}z}},
\]
with complex parameters $K_k$.

Since $\det\Phi(\lambda_k) =\mu^2 \det \tilde P(\lambda_k)\det \Phi_0 = 0$, there is a non-trivial eigenvector for each $k$. 
Solving this system of algebraic equations using Cramer's rule, one finds the $n$-soliton solution for the particular Minkowski
seed solution $\Phi_0$:
 \begin{equation}
\chi\left(\lambda,z,\bar{z}\right)=\left(\frac{K+\rm{i}\textit{z}}{K}\right)^{n}\frac{\mathbb A(\lambda)}{\mathbb A(\lambda=-1)}\quad\mbox{with}
\label{eq:solution_chi}
\end{equation}
\[\mathbb A(\lambda) = \left|\begin{array}{cccccc}
1 & -\lambda & \lambda^{2} & -\lambda^{3} & \cdots & \lambda^{2n}\\
1 & \alpha_{1}\lambda_{1} & \lambda_{1}^{2} & \alpha_{1}\lambda_{1}^{3} & \cdots & \lambda_{1}^{2n}\\
1 & \alpha_{2}\lambda_{2} & \lambda_{2}^{2} & \alpha_{2}\lambda_{2}^{3} & \cdots & \lambda_{2}^{2n}\\
1 & \alpha_{3}\lambda_{3} & \lambda_{3}^{2} & \alpha_{3}\lambda_{3}^{3} & \cdots & \lambda_{3}^{2n}\\
\vdots & \vdots & \vdots & \vdots & \ddots & \vdots\\
1 & \alpha_{2n}\lambda_{2n} & \lambda_{2n}^{2} & \alpha_{2n}\lambda_{2n}^{3} & \cdots & \lambda_{2n}^{2n}\end{array}\right|.\]
Here, the B{\"a}cklund parameters
\begin{equation}
 \alpha_k = -\frac{\chi(-\lambda_k)+\chi(\lambda_k)}{\chi(-\lambda_k)-\chi(\lambda_k)}
\end{equation}
were introduced.
Equation (\ref{eq:P3}) restricts the complex parameters to
 $K_i= \bar{K}_i$ and $\alpha_i \bar{\alpha}_i =1$ or $K_i= \bar{K}_k$ and $\alpha_i \bar{\alpha}_k =1$. 
The Ernst potential can be found by setting $\lambda=1$ in (\ref{eq:solution_chi}):
\begin{equation}
 f(\varrho,\zeta)=\chi(\lambda=1,\varrho,\zeta).\label{eq:solution_ernst}
\end{equation}
The solution (\ref{eq:solution_ernst}) of the Ernst equation (\ref{eq:ernst}) was derived by \cite{n79} by an 
iterated application of B\"acklund transformations to the Minkowski space (\cite{n80b}). 
Explicit formulae for the metric functions can be found in Appendix \ref{app:A}.

\subsection{Recovering the Ernst potential from its axis data}

Thus far, we discussed how to generate general solutions to Ernst's equation in the whole $(\rho,\zeta)$-plane. It is well-known, that
the axis values of the Ernst's potential uniquely determine the Ernst potential in the whole plane (cf. \cite{he81}).
The expression for $f$ found in the last section is a quotient of two polynomials in $\lambda_i$, and on the axis, this expression becomes a rational function in $\zeta$.
In this subsection, we will explore the correspondence between the 
so far (apart from the reality condition (\ref{eq:P3})) 
arbitrary B{\"a}cklund parameters $K_i,\alpha_i$ and a {\em given} Ernst potential 
on the upper $\zeta$-axis $\mathcal{A}^+$, denoted as
\begin{equation}
\label{f_rational_ansatz}
 f(\zeta) =\frac{\zeta^n+\sum_{j=1}^n a_j\zeta^{j-1}}{\zeta^n+\sum_{j=1}^n b_j\zeta^{j-1}} = \frac{Z(\zeta)}{N(\zeta)}.
\end{equation}
In the following, we will refer to Ernst potentials, that are rational functions on the axis as ``rational potentials''.
The formulae that will be presented in this subsection were found by Neugebauer.
They have also been discussed in \cite{m08} and found recently application in \cite{nh09}.

From (\ref{eq:linear_system_MN}) together with the definition of $A$ and $B$ as functions of the Ernst potential, 
one finds that the matrix $\Phi$ factorizes into a matrix 
depending solely on $\zeta$ times a matrix $D(K)$ depending only on $K$. 
The matrix $D(K)$ can be evaluated explicitly at the branch point 
$\mathcal{W}:=\left(\lambda=0,K=\zeta,\varrho=0\right)$, where $\lambda$ becomes single valued.
From (\ref{eq:solution_chi}) at $\lambda=1$  follows
that the parameters $K_i$ are the zeros of the polynomial\footnote{One has to consider the complex continuation of the real functions $Z(\zeta)$ and $N(\zeta)$.} 
$\frac{1}{2}\left[Z\left(\zeta\right)\bar{N}\left(\bar{\zeta}\right)+\bar{Z}\left(\bar{\zeta}\right)N\left(\zeta\right)\right]$. 
Combining these two results, one finds the mentioned
decomposition into $K$ and $\zeta$ dependent parts on the upper axis $\mathcal{A}^+$:

\[
\Phi(K,\zeta)=
\left(\begin{array}{cc}
\bar{f}(\zeta)  & 1\\
f(\zeta) & -1 \end{array}\right) D(K)\quad\mbox{and}\]
\[
D(K)=\frac{1}{2K^{n}}\left(\begin{array}{cc}
N\left(K\right)+\bar{N}\left(\bar{K}\right) & N\left(K\right)-\bar{N}\left(\bar{K}\right)\\
Z\left(K\right)-\bar{Z}\left(\bar{K}\right) & Z\left(K\right)+\bar{Z}\left(\bar{K}\right)\end{array}\right). 
\]

This allows us to show the following:

{\em
Let $f\left(\zeta\right)$ be a rational potential of the
form (\ref{f_rational_ansatz})
with arbitrary parameters $a_{i}$, $b_{i}$.
Let furthermore $K_{i}\neq K_{j}$ and $\alpha_{i}\neq\alpha_{j}$
$\forall i\neq j$. For any set of parameters $\left\{ a_{i},b_{i}\right\} $
there is exactly one set of parameters $\left\{ K_{i},\alpha_{i}\right\} $,
such that the B{\"a}cklund transformation $\mathcal{B}$ acting on the Minkowski seed solution results in an Ernst potential $f$ 
\[\mathcal{B}\left[f_{0}=1,K_{i},\alpha_{i}\right]=f\]
that becomes the given rational potential on the axis.
The B{\"a}cklund parameters can be explicitly calculated by evaluating
the expressions
\begin{equation}
\frac{1}{2}\left[Z\left(\zeta\right)\bar{N}\left(\bar{\zeta}\right)+\bar{Z}\left(\bar{\zeta}\right)N\left(\zeta\right)\right]=\prod_{i=1}^{2n}\left(\zeta-K_{i}\right),\label{eq:K_i}\end{equation}
\begin{equation}
\frac{\bar{N}\left(\bar{K}_{i}\right)}{N\left(K_{i}\right)}=-\frac{\bar{Z}\left(\bar{K}_{i}\right)}{Z\left(K_{i}\right)}=\alpha_{i}.\label{eq:alpha}\end{equation}
}
Proof:\newline
Evaluation of\begin{eqnarray*}
\Phi\left(K_{i}\right)\left(\Phi_{0}\right)^{-1}\left(\begin{array}{c}
-\alpha_{i}\\
1\end{array}\right) & = & 0\end{eqnarray*}
leads to (\ref{eq:alpha}).
This overdetermined system of equations for the $\alpha_{i}$
has solutions, iff $Z\left(K_{i}\right)\bar{N}\left(\bar{K}_{i}\right)+\bar{Z}\left(\bar{K}_{i}\right)N\left(K_{i}\right)=0$.
Another way of proving this fact is by simply evaluating the 
expressions (\ref{eq:alpha}) and (\ref{eq:K_i}) using the representation of $f$ in form of determinants (e.g. (\ref{eq:solution_chi})) 
at $\lambda=1$, and verifying that they are fulfilled.
\newline
{\em q.e.d.}

The general algorithm to determine $f\left(\varrho,\zeta\right)$
from $f\left(\varrho=0,\zeta\right)$ works as follows: First
of all, one has to calculate the $K_{i}$ by evaluating (\ref{eq:K_i}).
The fundamental lemma of algebra guarantees existence and uniqueness.
Then one can easily find the $\alpha_{i}$ by evaluating (\ref{eq:alpha}). 
The equations (\ref{eq:K_i}) and (\ref{eq:alpha}) are universal in
the sense that one does not need to know beforehand whether the
$K_{i}$ or the $\alpha_{i}$ are complex or real, nor does the
order of the polynomial matter. Equation (\ref{eq:alpha}) allows the calculation of $\alpha_i$
even if $K_i$ is a zero of $N$ or of $Z$.
The result can be easily generalized by using Bernoulli's rule for coinciding zeros ($K_i=K_j$ for $i\neq j$).  
The $\alpha_{i}$ have to be treated as usual functions of $K_{i}$. 

\section[Constructing rational Ernst potential on the rotation axis]{Different approaches of constructing rational Ernst potentials on the rotation axis}

\subsection{Reflectional symmetric Ernst potentials}
In Newtonian theory one can show that rotating fluid bodies in equilibrium always have reflectional symmetry (cf. \cite{l33}). The same symmetry is to be expected in Einsteinian theory (cf. \cite{l92}) and we assume it to exist for our purpose.
Therefore we study the influence of reflectional symmetry before constructing rational potentials.
As shown in \cite{mn95} and \cite{k95}, 
this kind of symmetry of the metric is uniquely characterized by a simple relation for the Ernst potential $f$ 
on the upper part $\mathcal{A}^+$ of the $\zeta$-axis:
\begin{equation}\label{symmetrycond}
 f(\zeta)\overline{ f(-\zeta)} = 1.
\end{equation}
Condition (\ref{symmetrycond}) is fulfilled for all $\zeta$, iff $N(\zeta) = \overline{Z(-\zeta)}$. As shown in the proof of Theorem 2 of \cite{k95}, this is equivalent to the following conditions for the coefficients:
\begin{equation}\label{condition}
b_{n}=-\bar a_n,\quad b_{n-1} = \bar a_{n-1}\quad\dots\quad b_1 = \left\{ \\
\begin{array}{cc}
\bar a_1, & n ~ \mathrm{even}\\
-\bar a_1, & n ~ \mathrm{odd}
\end{array}
\right. .
\end{equation}

So we have reduced the $2n$ complex parameters to $2n$ real ones.

\subsection{Construction from multipole moments}\label{constructionfrommn}
As a first method of generating a rational Ernst potential $f(\zeta)$ on the rotation axis, we refer to \cite{mr98}. They start from the expansion of the function $X(\zeta)$ at $\zeta = \infty$:
\begin{equation}\label{xi}
 X(\zeta) \equiv \frac{1-f(\zeta)}{1+f(\zeta)}=\sum\limits_{j=0}^\infty m_j\zeta^{-j-1},
\end{equation}
where the coefficients $m_j$ determine the Geroch-Hansen multipole moments $P_j$ (the explicit relations between $m_j$ and $P_j$ are given in \cite{fhp89}). After plugging in the ansatz (\ref{f_rational_ansatz}) and evaluating the algebraic set of equations, which results from comparing the coefficients of the powers of $\zeta$ up to $n = 2 N$ in (\ref{xi}), one ends up with
\begin{equation}
 Z(\zeta) = \frac{\left\vert\begin{array}{cccc}
\zeta^n-\sum_{j=0}^{n-1}m_j\zeta^{n-1-j} & m_n & \dots & m_{2n-1}\\
\zeta^{n-1}-\sum_{j=0}^{n-2}m_j\zeta^{n-2-j} & m_{n-1} & \dots & m_{2n-2}\\
\vdots & \vdots & \ddots & \vdots\\
\zeta-m_0 & m_1 & \dots & m_{n}\\
1 & m_0 & \dots & m_{n-1}
\end{array}\right\vert}{\left\vert\begin{array}{cccc}
m_{n-1} & m_n & \dots & m_{2n-2}\\
m_{n-2} & m_{n-1} & \dots & m_{2n-3}\\
\vdots & \vdots & \ddots & \vdots\\
m_1 & m_2 & \dots & m_{n}\\
m_0 & m_1 & \dots & m_{n-1}
\end{array}\right\vert}
\end{equation}
and an analogue formula for $N(\zeta)$ by changing the sign ``$-$'' to ``$+$'' in the first column of the numerator. From this formula one can see that condition (\ref{condition}) is automatically fulfilled when $m_j$ is real for even $j$ and purely imaginary otherwise, which is another well-known formulation of reflection symmetry (see \cite{k95}).

Since only far-field information is employed by use of the multipole moments, it is guaranteed that the $f(\zeta)$ constructed in this way will coincide well with the solution from which the multipole moments were taken in a neighbourhood of infinity.

\subsection{Fitting a given (non-rational) Ernst potential}

Another method is constructing a rational potential $f(\zeta)$ from a given function $g(\zeta)$ on the rotation axis,
 which is especially useful in a numerical context. There, points $\zeta_i$ on the axis are selected where $f(\zeta)$ shall 
exactly coincide with $g(\zeta)$. If the points $\zeta_i$ are suitably chosen, we
find a good approximation $f$ of $g$ on the whole axis. The same ansatz (\ref{f_rational_ansatz}) is employed.

A major advantage of this method is that $f(\zeta)$ and $g(\zeta)$ will also coincide near the star's surface, if one or some of the $\zeta_i$'s are situated there.

Rational potentials found this way do not automatically have the expected behaviour at large distances (all multipole moments and especially mass $M=m_0=P_0$ and angular momentum $J=-$i$m_1=-$i$P_1$ of the analytical solution may vary from the given numerical ones). To guarantee the right asymptotic behaviour, we treated the coefficients in (\ref{f_rational_ansatz}) as functions of $m_n$ at fixed $M$ and $J$. The $m_n$ were varied until $f(\zeta)$ and $g(\zeta)$ coincided at the gridpoints $\zeta_i$.

\section{Results}
After constructing a rational Ernst potential one has to calculate the B\"acklund parameters according to formulae (\ref{eq:K_i}) and (\ref{eq:alpha}). 
For polynomials 
in numerator and denominator of degree $n$ there are $2n$ real parameters.
The analytical approximate solution (often referred to as B\"acklund solution, e.g. as presented in formula (\ref{eq:solution_chi})) is a $2n$-fold B\"acklund transformation acting on the Minkowski seed solution.

In the following we present solutions with $n=1,2,3$ and 4 to show the increasing accuracy of the approximation when taking more and more parameters into account.

For our numerical calculations we used the AKM programme (\cite{akm03}), where different coordinates had been introduced. We have chosen to compare our results in Weyl-Lewis-Papaetrou coordinates and have taken the necessary transformations into account.

\subsection{Comparison of metric functions}
\begin{figure}[ht]
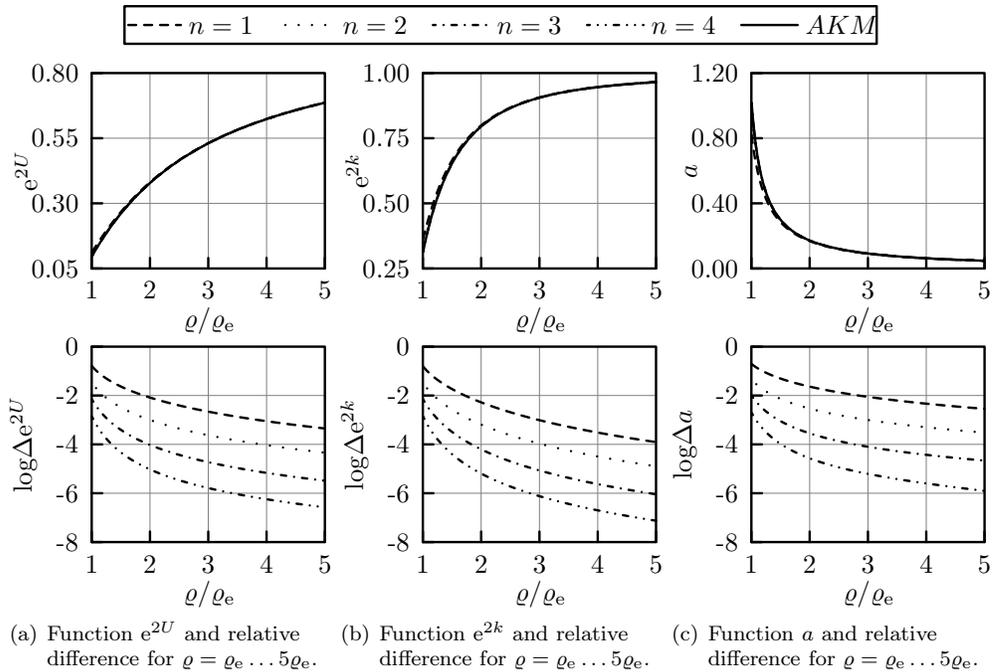

 \centering
 \includegraphics{figure0.eps}\\
\subfigure[Function e$^{2U}$ and relative \newline difference for $\varrho=\varrho_{\mathrm e}\dots 5\varrho_{\mathrm e}$.]{\includegraphics{figure1a.eps}}\hfill
 \subfigure[Function e$^{2k}$ and relative \newline difference for $\varrho=\varrho_{\mathrm e}\dots 5\varrho_{\mathrm e}$.]{\includegraphics{figure1b.eps}}\hfill
 \subfigure[Function $a$ and relative \newline difference for $\varrho=\varrho_{\mathrm e}\dots 5\varrho_{\mathrm e}$.]{\includegraphics{figure1c.eps}}
\caption{\small Metric potentials as functions of coordinate radius $\varrho$ in the equatorial plane (homogeneous equation of state and fitted rational potential on the axis with fixed $M$ and $J$). }\label{metrichomogen}
\end{figure}

As a first result we have calculated the metric functions and the relative difference between analytical and numerical values (see figure \ref{metrichomogen}). For the numerical calculations one has to choose an equation of state. 
In this section we decided to model our neutron star to be composed of matter of constant energy density. The star we have chosen to treat as an example has a ratio of polar to equatorial radius of 0.7 (this ratio is always given in the coordinates used in \cite{akm03}) and a normalized central energy density of 1 in order to compare the results with \cite{nozawa} and \cite{akm03}.

These approximate solutions were constructed by fitting the given numerical Ernst potential on the axis, holding $M$ and $J$ fixed. For $n=1$ (i.e. a twofold Bäcklund transformation) this already determines the two 
available real parameters so that no additional gridpoint can be choosen. This is equivalent to approximating the neutron star's exterior metric by a Kerr solution with the same mass and angular momentum. We expect therefore the approximation to coincide well with the numerically calculated potentials only near infinity, as we already explained in the end of section \ref{constructionfrommn}, and to differ most near the star's surface.

In a $2n$-fold Bäcklund transformation with $n > 1$ one has to fix more parameters. 
In fitting the Ernst potential on the axis, our algorithm chooses a new gridpoint 
where - in the previously calculated $(2n-1)$-fold transformation - the approximation and the numerical potential differ most. 
For a fourfold Bäcklund transformation, this is the star's surface, while for higher transformations the additional 
gridpoints are usually well spaced on the axis, which shows the efficiency of our algorithm.

As one may observe e.g. in figure \ref{metrichomogen}, one gains about one order of magnitude in 
accuracy by going from a $n$-fold to a $(n+1)$-fold Bäcklund transformation, i.e. with two parameters more.

\subsection{Comparison of physical properties}
As we approximate the exterior vacuum region of a numerically calculated spacetime it is necessary to define what physical properties mean in our context. Of course one can simply read off mass $M$ and angular momentum $J$ from the far-field behaviour of the metric, as well as higher multipoles. This is sufficient if one only wants to analyse far-field behaviour. 

\begin{figure}[ht]
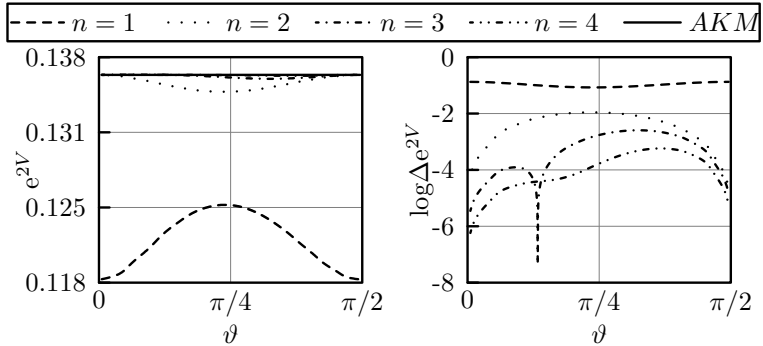

 \centering
\includegraphics{figure0.eps}
 \includegraphics{figure2.eps}
\caption{\small Potential e$^{2V}$ and relative difference as a function of $\vartheta:=\arctan(\varrho/\zeta)$. e$^{2V}$ coincides at pole and equator (for $n>1$), because we obtained the analytical vacuum metric by fitting the numerical Ernst potential at the rotation axis (and the first gridpoint is the pole $\vartheta=0$). }\label{e2Vhomogen}
\end{figure}

A more intriguing question is how to determine the star's surface in the appro\-ximated solution. Using the co-rotating potential $V \equiv U^\prime$ defined by (see e.g. \cite{maknp08})
\[\mathrm{e}^{2U^\prime} := \mathrm{e}^{2U}\left[(1+\Omega a)^2-\Omega^2\varrho^2\mathrm{e}^{-4U}\right],\]
the star's boundary is given by $V = V_0 =$ const. However, given only the exterior metric (and not the rotation speed $\Omega$ of the star), it is not possible to construct the co-rotating potentials directly. Our approach to deal with this problem is the following:
Since on the rotation axis $\varrho = 0$ the co-rotating potentials coincide with the non-rotating ones, we read off the approximation of $V_0$ from the approximated potential $U$ at the north pole $V_0=U^\prime(\varrho=0,~\zeta=\zeta_{\rm p})=U(\varrho=0,~\zeta=\zeta_{\rm p})$ 
and use the constancy of $V_0$ to calculate the rotation speed $\Omega$ at the equator
\begin{equation}\label{eq:V_0}
 \mathrm{e}^{2V_0}=\mathrm{e}^{2U^\prime}(\varrho=\varrho_{\mathrm e},\zeta=0).
\end{equation}
The values $\varrho_{\rm e}$ and $\zeta_{\rm p}$ are supplied by the numerical solution.
Once $\Omega$ is known we can evaluate $V \equiv U^\prime$ in the whole vacuum region and $V=V_0$ is valid at the pole and equator by construction. Of course, the so constructed $V$ will not be constant over the original (numerical) star's surface, and also $V_0$ will differ from the value in the numerically calculated solution if $U$ does not coincide at the pole. This can be seen in the change of behaviour from $n=1$ to $n=2$, since for $n>1$ a fitting gridpoint is located at the star's pole.

To illustrate this behaviour we plotted the potential e$^{2V}$ along the star's surface and the relative difference between the $2n$-fold B\"acklund solution and the AKM solution in figure \ref{e2Vhomogen}.

\begin{figure}[ht]
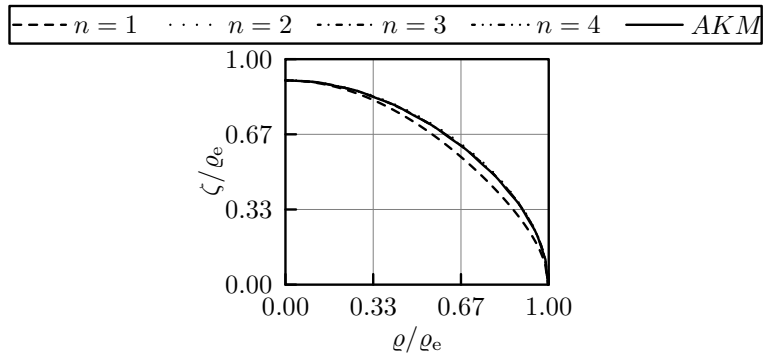

\centering
\includegraphics{figure0.eps}
 \includegraphics{figure3.eps}
\caption{\small Area e$^{2U^\prime}=\mathrm{e}^{2U}\left[(1+\Omega {a})^2-\Omega^2\varrho^2\mathrm{e}^{-4U}\right]\equiv \mathrm{e}^{2V_0}$ in comparison with the star's surface.}\label{surfhomogen}
\end{figure}

To remedy also this non-constantness, we simply \textit{define} the star's surface in our approximation as the region where e$^{2U^\prime} =$ e$^{2V_0}$. One can see in figure \ref{surfhomogen} that this area is almost identical with the star's surface for solutions generated by $2n$-fold B\"acklund transformations with $n>1$.
This way of interpreting the analytical solution has the advantage that one only needs two surface points (the pole $\zeta_{\rm p}$ and the equator $\varrho_{\rm e}$) and does not need to worry about the whole surface function $\varrho_s(\zeta)$.

Finally we calculated the following physical quantities in table \ref{tab:1} (formulae can be found in \cite{nozawa} or \cite{akm03}):
\begin{center}
\begin{tabular}{ll}
 $R_{\rm circ}$ & equatorial circumferential radius\\
 $Z_{\rm p}$ & polar redshift\\
 $Z^{\rm f}_{\rm eq}$ & equatorial redshift in the forward direction\\
 $Z^{\rm b}_{\rm eq}$ & equatorial redshift in the backward direction.
 \end{tabular}
\end{center}

\begin{table*}[htb]
\begin{center}\small
\caption{\small Detailed comparison of physical quantities (homogeneous equation of state ($\mu =\mu_0$) and fitted rational potential on the axis with fixed $M$ and $J$). Here, $\tilde p_{\rm c} = p_{\rm c} /\mu_0,~\tilde \Omega = \Omega/\mu_0^{1/2},~\tilde M = M\mu_0^{1/2},~\tilde R_{\rm circ} =R_{\rm circ}\,\mu_0^{1/2}$ and $\tilde J = J\mu_0$ are normalized values of the physical quantities. The column labelled AKM shows numerically calculated values, whereas BT[$n$] refers to values of a 2$n$-fold B\"acklund solution. $\Delta$BT[$n$] labels the relative difference between BT[$n$] and AKM.}\label{tab:1}
\begin{tabular}{cccccccccc}\br
Model  &\hspace{-2.5mm}  AKM    &\hspace{-2.5mm}  BT[1]   &\hspace{-2.5mm}   BT[2]   &\hspace{-2.5mm}   BT[3]   &\hspace{-2.5mm}   BT[4]  &\hspace{-2.5mm}    $\Delta$BT[1]   &\hspace{-2.5mm} $\Delta$BT[2]  &\hspace{-2.5mm}  $\Delta$BT[3]   &\hspace{-2.5mm}  $\Delta$BT[4] \\\mr
$\tilde p_{\rm c} 	$	&\hspace{-2.5mm} 1.00\\
$r_{\rm p}/r_{\rm e}$	&\hspace{-2.5mm} 0.700 \\
$\tilde\Omega 	$	&\hspace{-2.5mm} 1.41 &\hspace{-2.5mm}  1.58 &\hspace{-2.5mm}  1.42 &\hspace{-2.5mm}  1.41 &\hspace{-2.5mm}  1.41 &\hspace{-2.5mm}  1.2e-01 &\hspace{-2.5mm}  7.4e-03 &\hspace{-2.5mm}  1.4e-03 &\hspace{-2.5mm}  2.9e-04\\
$\tilde M 	$		&\hspace{-2.5mm} 0.136 &\hspace{-2.5mm}  0.136 &\hspace{-2.5mm}  0.136 &\hspace{-2.5mm}  0.136 &\hspace{-2.5mm}  0.136 \\
$\tilde J 	$		&\hspace{-2.5mm} 0.0141 &\hspace{-2.5mm}  0.0141 &\hspace{-2.5mm}  0.0141&\hspace{-2.5mm}  0.0141 &\hspace{-2.5mm}  0.0141\\
$\tilde R_{\rm circ}$ 		&\hspace{-2.5mm} 0.345 &\hspace{-2.5mm}  0.337 &\hspace{-2.5mm}  0.344 &\hspace{-2.5mm}  0.345 &\hspace{-2.5mm}  0.345 &\hspace{-2.5mm}  2.4e-02 &\hspace{-2.5mm}  4.6e-03 &\hspace{-2.5mm}  9.3e-04 &\hspace{-2.5mm}  1.9e-04\\
$Z_{\rm p} 	$	&\hspace{-2.5mm} 1.71 &\hspace{-2.5mm}  1.91 &\hspace{-2.5mm}  1.71 &\hspace{-2.5mm}  1.71 &\hspace{-2.5mm}  1.71 &\hspace{-2.5mm} 1.2e-01\\
$Z^{\rm f}_{\rm eq} $	&\hspace{-2.5mm} -0.163 &\hspace{-2.5mm}  -0.295  &\hspace{-2.5mm} -0.170 &\hspace{-2.5mm}  -0.164 &\hspace{-2.5mm}  -0.163 &\hspace{-2.5mm}  8.1e-01 &\hspace{-2.5mm}  4.3e-02 &\hspace{-2.5mm}  8.5e-03 &\hspace{-2.5mm}  1.8e-03\\
$Z^{\rm b}_{\rm eq}$	&\hspace{-2.5mm} 11.4 &\hspace{-2.5mm}  11.7&\hspace{-2.5mm}  11.1 &\hspace{-2.5mm}  11.3 &\hspace{-2.5mm}  11.3 &\hspace{-2.5mm}  2.6e-02 &\hspace{-2.5mm}  2.8e-02 &\hspace{-2.5mm}  5.7e-03 &\hspace{-2.5mm}  1.1e-03\\\br
\end{tabular}
\end{center}
\end{table*}
\normalsize

Analogous results for different equations of state can be found in Appendix \ref{app:B}.

\subsection{Critical points and singularities}
A closer look at the spectral parameter $\lambda=\sqrt{(K-\mathrm i \bar z)/(K+\mathrm i z)}$, where the analogously defined $\lambda_i$ are  part of the formulae for the metric functions, indicates some critical points, because the square root has a branch cut which we took along the negative real axis. One easily sees that the points
\[ z_{k1} := \mathrm i \bar K_k= \Im{K_k} +\mathrm i \Re{K_k},\]
where the radicand of $\lambda_k$ vanishes and 
\[ z_{k2} := \mathrm i K_k= -\Im{K_k} +\mathrm i \Re{K_k},\]
where the radicand of $\lambda_k$ goes to infinity can become problematic. For example in the case of a 4-fold B\"acklund transformation with 4 complex $K_i$ ($K_1,~K_2=\bar K_1,~K_3 = -\bar K_1$ and $K_4 = - K_1$) the metric function e$^{2U}$ has a jump along $g:\{\varrho\in [0,\Im K_1],\zeta = \Re K_1\}.$

\begin{figure}[ht]
 \centering
 \includegraphics{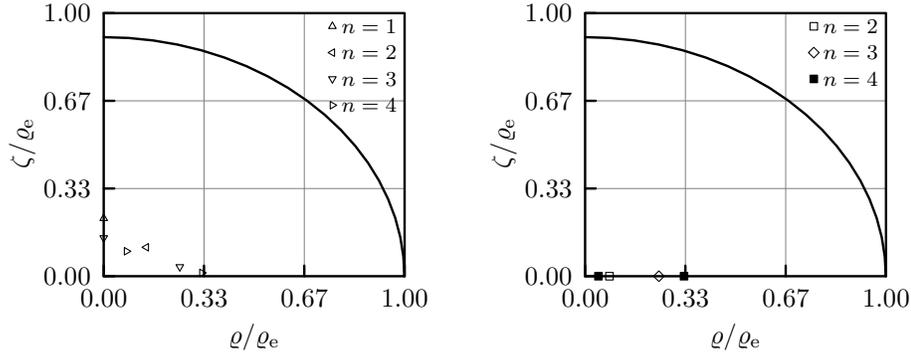}
\caption{\small Star's surface (solid line), critical points (triangles) and singularities (boxes) of the B\"acklund solutions. The exterior solution is well behaved, because the difficult points are inside the star and the singularities are situated near the critical points.}\label{singularities}
\end{figure}

As the formulae for the calculation of the metric functions contain determinants in the denominator, even singularities of the analytic solution may occur. Therefore we have numerically calculated the zeros of the denominator (see figure \ref{singularities}) and roughly speaking they usually appear near or between the $K_i$. 

All these problems are controllable for our purpose. As we search for an analytical vacuum solution for the exterior of a neutron star, this is well behaved as long as the critical points and singularities are located in the interior region of the star. To decide whether our solution is applicable or not, we simply added a plot of the star's surface with the critical points to the output of our approximation programme.

\subsection{Numerical methods}

To calculate the approximation, we augmented the AKM programme (\cite{akm03}) to incorporate the formulae given above. Two main tasks had to be done: Calculation of determinants (e.g. (\ref{f}),(\ref{a}) and (\ref{e2k})) and root finding for polynomials (\ref{eq:K_i}).

To calculate the determinants, Gaussian forward elimination was used to convert the matrix to triangular form, whence the determinant is given by multiplication of the diagonal elements. For our purposes, this method was found to be numerically stable by using partial pivoting, i.e. transposing columns until the largest element in each row was on the diagonal.

Polynomial root finding was accomplished for $n=2$ and $n=3$ by direct solving the quadratic or cubic equation, while for $n>3$ we determined the roots as the eigenvalues of the companion matrix. However, Gaussian elimination was not sufficient in this case and we had to resort to a QR decomposition of the companion matrix (adapted from the GSL (cf. \cite{gsl})).

At last, to find the singularities, we had to search for zeros of the denominator in (\ref{f}). Since the determinant $\mathbb A^+$, viewed as function in the complex $(\rho,\zeta)$-plane, has branch cuts where it is not even continous, the obvious Newton's method fails spectacularly. The Nelder-Mead simplex algorithm (\cite{nm65}) does work for non-continuous functions and is well behaved for this problem. We start with a number of random trial points in the $(\rho,\zeta)$-plane, from which this algorithm searches local minima of the absolute value of the determinant. Afterwards we check if these minima really are zeros.

\subsection{Conclusion}\label{conclusion}
It is remarkable that the whole exterior region of a stationary and axisymmetric neutron star can be expressed with an accuracy greater than 99\% everywhere using only 6 to 10 real parameters ($\varrho_{\rm e}$ and $\zeta_{\rm p}$ to determine the star's surface and 4 to 8 multipole moments, $m_n$ or values of Ernst's potential on the rotation axis). As there are many ways to construct a rational potential on the axis one may choose whether near or far-field information should be considered. From this point of view, there is no ``best approximation'', but rather it can be adjusted for specific needs.

In each case one has to be aware of the singularities and as long as they are located in the interior of the star, there is no problem.

\begin{appendix}

\section{Explicit formulae for the determination of metric functions}\label{app:A}
An alternative formula (obtainable from (\ref{eq:solution_chi}) by performing the limit $\lambda\rightarrow 1$, some manipulation of the determinants and inserting the definition of $r_i$)  for the calculation of Ernst's potential is
\begin{equation}\label{f}
 f(\varrho,\zeta)= \frac{\mathbb A^-}{\mathbb A^+}, \quad\mathbb A^\mp = \left|
 \begin{array}{ccccc}
  1 & \mp 1 & 0 & \dots & 0\\
  K_1^n & K_1^{n-1}\alpha_1r_1 & K_1^{n-1} & \dots & 1\\ 
  K_2^n & K_2^{n-1}\alpha_2r_2 & K_2^{n-1} & \dots & 1\\
  \vdots & \vdots & \vdots & \ddots & \vdots\\
  K_{2n}^n & K_{2n}^{n-1}\alpha_{2n}r_{2n} & K_{2n}^{n-1} & \dots & 1
 \end{array}
\right| ,
\end{equation}
with $r_i = \lambda_i(K_i+\mathrm{i}z)$.
Of course, e$^{2U}$ is the real part of $f$ and one could obtain $a$ from its imaginary part $b$ via (\ref{eq:b}).
A straightforward calculation leads to the explicit formula
\begin{equation}\label{a}
 (a-a_0)\mathrm{e}^{2U} = 2n\varrho+2\Re\left(\frac{\mathbb B}{\mathbb A^+}\right),
\end{equation} 
\[\mathbb B =
\left|
 \begin{array}{cccccc}
  -nz & -\varrho-\left(n-1\right)z & -\mathrm{i}& -\mathrm{i} &  \dots & 0\\
  K_1^{n} & \alpha_1r_1K_1^{n-1} & K_1^{n-1} & \alpha_1r_1K_1^{n-2} & \dots & 1\\ 
  K_2^{n} & \alpha_2r_2K_2^{n-1} & K_2^{n-1} & \alpha_2r_2K_2^{n-2} & \dots & 1\\ 
  \vdots & \vdots & \vdots & \vdots & \ddots &\vdots \\
  K_{2n}^{n} & \alpha_{2n}r_{2n}K_{2n}^{n-1} & K_{2n}^{n-1} &  \alpha_{2n}r_{2n}K_{2n}^{n-2} & \dots &1\\ 
 \end{array}
\right|,\]
where the integration is already done. The only thing left to be done is the determination of the constant $a_0$, which can be calculated by demanding that $a\rightarrow 0$ as $\varrho \rightarrow\infty$. 

Furthermore the last metric function e$^{2k}$ could be calculated via line integration of (\ref{eq:k}). By reformulating the linear problem (\ref{eq:linear_system_MN}) and some manipulation of determinants (cf. \cite{knm91}, \cite {k80}) one ends up with the following formula:
\begin{equation}\label{e2k}
 \mathrm{e}^{2(k-k_0)} = \prod\limits_{j=1}^{2n}\frac{1}{r_j(K_j +\mathrm{i}z)}~\mathbb D_1 \mathbb D_2,
\end{equation}
\[\mathbb D_1 =  \left|
\begin{array}{cccc}
 (K_1+\mathrm{i}z)^n & \alpha_1 r_1 K_1^{n-1} & \dots &\alpha_1r_1\\
 (K_2+\mathrm{i}z)^n & \alpha_2 r_2 K_2^{n-1} & \dots &\alpha_2r_2\\
  \vdots & \vdots & \ddots & \vdots\\
 (K_{2n}+\mathrm{i}z)^n & \alpha_{2n}r_{2n}K_{2n}^{n-1}  & \dots &\alpha_{2n}r_{2n}
\end{array}\right|\]
and $\mathbb D_2$ is the same as $\mathbb D_1$ with $\alpha_i$ replaced by $1/\alpha_i$. Finally the integration constant $k_0$ can be evaluated by demanding that e$^{2k}\rightarrow 1$ as $\varrho\rightarrow\infty$.

\section[Results for different equations of state]{Results for different equations of state and approximation methods}\label{app:B}

In order to avoid a flood of data we do not present the results for three different equations of state (strange quark, constant energy density and polytropic) with three different approximation methods (via moments $m_n$, fitting a nonrational potential and fitting a nonrational potential with fixed $M$ and $J$). The following figures and tables provide one example of each method, but not necessarily the best one for the given equation of state. 

\subsection{Polytropic stars approximated by moments $m_n$}

As an example with polytropic equation of state (polytropic exponent $\Gamma =2$, polytropic constant $K$) we have chosen $\tilde{\mu}_{\rm c}=1$ and $r_{\rm p}/r_{\rm e}=0.834$ in order to compare the values for the physical quantities with \cite{nozawa} and \cite{akm03}. 

\begin{figure}[ht]
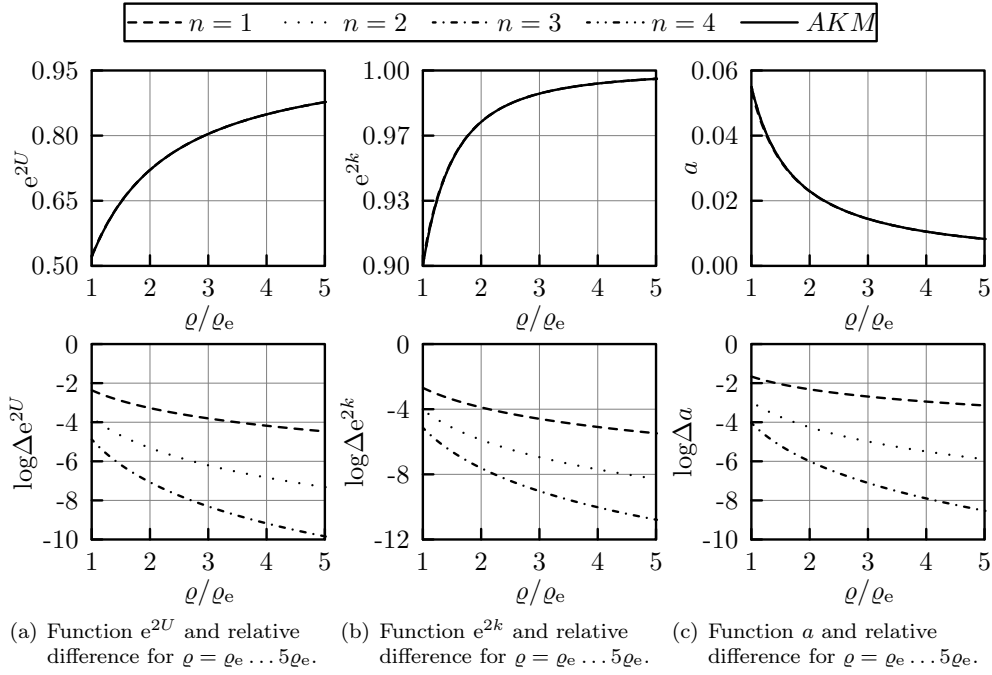

 \centering
\includegraphics{figure0.eps}\\
 \subfigure[Function e$^{2U}$ and relative\newline  difference for $\varrho=\varrho_{\mathrm e}\dots 5\varrho_{\mathrm e}$.]{\includegraphics{figureB1a.eps}}\hfill
 \subfigure[Function e$^{2k}$ and relative\newline difference for $\varrho=\varrho_{\mathrm e}\dots 5\varrho_{\mathrm e}$.]{\includegraphics{figureB1b.eps}}\hfill
 \subfigure[Function $a$ and relative\newline difference for $\varrho=\varrho_{\mathrm e}\dots 5\varrho_{\mathrm e}$.]{\includegraphics{figureB1c.eps}}
\caption{\small Metric potentials as functions of coordinate radius $\varrho$ in the equatorial plane (polytropic equation of state and rational potential via moments $m_n$ on the axis). }\label{metricpoly}
\end{figure}

A first look at figure \ref{metricpoly} shows, that all approximated functions coincide with errors less than 1\%.
The eightfold B\"acklund solution is missing from those data. Since critical points and singularities appeared outside the star, we had to discard this solution. As these approximated solutions are generated only by far-field information we observed quite generally that critical points appear more often outside the star.

As this approximation method does not include any information about the star's surface, at first sight one could think that e$^{2V}$ is not approximated very well. The graphs (see figure \ref{e2Vpoly}) however show quite a good approximation even in this case (especially for $n>1$).  Remembering the definition of $V_0$ and $\Omega$ (cf. (\ref{eq:V_0})) it is clear that the values of e$^{2V}$ for $\vartheta =0$ and $\vartheta = \pi/2$ coincide.

\begin{figure}[ht]
 \centering
\includegraphics{figure0.eps}
 \includegraphics{figureB2.eps}
\caption{\small Potential e$^{2V}$ as a function of $\vartheta:=\arctan(\varrho/\zeta)$.  }\label{e2Vpoly}
\end{figure}

Figure \ref{surfpoly} confirms that the real star's surface coincides well with the one determined in the approximation by e$^{2U^\prime}\equiv\mathrm{e}^{2V_0}$.

\begin{figure}[ht]
\centering
\includegraphics{figure0.eps}
 \includegraphics{figureB3.eps}
\caption{\small Area e$^{2U^\prime}=\mathrm{e}^{2U}\left[(1+\Omega {a})^2-\Omega^2\varrho^2\mathrm{e}^{-4U}\right]\equiv \mathrm{e}^{2V_0}$ in comparison with the star's surface. }\label{surfpoly}
\end{figure}
Finally, we calculated some physical quantities in table \ref{tab:poly}.

\begin{table*}[htb]
\begin{center}\small
\caption{\small Detailed comparison of physical quantities (polytropic equation of state ($\mu = p+\sqrt{p/K}$) and rational potential via $m_n$). Here, $\tilde \mu_{\rm c} = \mu_{\rm c} K,~\tilde \Omega = \Omega K^{1/2},~\tilde M = M/K^{1/2},~\tilde R_{\rm circ} =R_{\rm circ}/K^{1/2}$ and $\tilde J = J/K$ are normalized values of the physical quantities. The column labelled AKM shows numerically calculated values, whereas BT[$n$] refers to values of a 2$n$-fold B\"acklund solution. $\Delta$BT[$n$] labels the relative difference between BT[$n$] and AKM.}\label{tab:poly}
\begin{tabular}{cccccccc}\br
Model  &  AKM    &  BT[1]   &   BT[2]   &   BT[3]   &     $\Delta$BT[1]   & $\Delta$BT[2]  &  $\Delta$BT[3] \\\mr
$\tilde \mu_{\rm c}$	& 1.00 \\
$r_{\rm p}/r_{\rm e}$	& 0.834 \\
$\tilde \Omega 	$	& 0.400&  0.425&  0.400&  0.401&  6.1e-02 &  1.8e-03 &  2.4e-04 \\
$\tilde M 	$		& 0.161&  0.161&  0.161&  0.161\\
$\tilde J 	$		& 9.49e-03 &  9.49e-03 &  9.49e-03 &  9.49e-03 \\
$\tilde R_{\rm circ}$ 		& 0.679 &  0.678 &  0.679 &  0.679 &  2.1e-03 &  8.2e-05 &  5.8e-06\\
$Z_{\rm p} 	$	& 0.458 &  0.467 &  0.458 &  0.458 &  2.0e-02 &  1.0e-03 &  8.0e-05\\
$Z^{\rm f}_{\rm eq} $	& -0.0601&  -0.0845 & -0.0594&  -0.0602&  4.1e-01 &  1.2e-02 &  1.6e-03\\
$Z^{\rm b}_{\rm eq}$	& 1.04 &  1.09 &  1.04 &  1.04 &  4.4e-02 &  1.8e-03 &  1.8e-04\\\br
\end{tabular}
\end{center}
\end{table*}
\normalsize

\subsection{Strange quark stars approximated by fitting Ernst's potential at the rotation axis}

\begin{figure}[ht]
 \centering
\includegraphics{figure0.eps}\\
 \subfigure[Function e$^{2U}$ and relative\newline difference for $\varrho=\varrho_{\mathrm e}\dots 5\varrho_{\mathrm e}$.]{\includegraphics{figureB4a.eps}}\hfill
 \subfigure[Function e$^{2k}$ and relative\newline difference for $\varrho=\varrho_{\mathrm e}\dots 5\varrho_{\mathrm e}$.]{\includegraphics{figureB4b.eps}}\hfill
 \subfigure[Function $a$ and relative\newline difference for $\varrho=\varrho_{\mathrm e}\dots 5\varrho_{\mathrm e}$.]{\includegraphics{figureB4c.eps}}
 \caption{\small Metric potentials as functions of coordinate radius $\varrho$ in the equatorial plane (strange quark equation of state and fitted rational potential on the axis). }\label{metricstrange}
\end{figure}

As an example with strange quark equation of state (linear relation between energy density and pressure, MIT-Bag constant $B$) we have chosen $\tilde{p}_{\rm c}=2$ and $r_{\rm p}/r_{\rm e}=0.5$ in order to compare the values for the physical quantities with \cite{akm03}. As the star is quite relativistic, figure \ref{metricstrange} shows that all approximated functions do not reach the accuracy of the other examples.

\begin{figure}[ht]
\centering
\includegraphics{figure0.eps}
 \includegraphics{figureB5.eps}
\caption{\small Potential e$^{2V}$ as a function of $\vartheta:=\arctan(\varrho/\zeta)$. e$^{2V}$ coincides at pole and equator, because we obtained the analytical vacuum metric by fitting the numerical Ernst's potential at the rotation axis (and the first gridpoint is the pole $\vartheta=0$). }\label{e2Vstrange}
\end{figure}

The next figure shows the area e$^{2U^\prime}\equiv\mathrm{e}^{2V_0}$. The fact that the curve for the $n=1$ B\"acklund solution does not end at the equatorial radius, might be confusing at the first sight (especially when in figure \ref{e2Vstrange} the error of e$^{2V}$ vanishes there). As the surface area e$^{2U^\prime}$ is not necessarily a monotonic function, there can be two intersection curves that need not be connected. In figure \ref{surfstrange} this is the case, where one intersection curve ends at $\varrho\approx0.67\varrho_{\rm e}$ and the other starts at $\varrho=\varrho_{\rm e}$.

\begin{figure}[ht]
\centering
\includegraphics{figure0.eps}
 \includegraphics{figureB6.eps}
\caption{\small Area e$^{2U^\prime}=\mathrm{e}^{2U}\left[(1+\Omega {a})^2-\Omega^2\varrho^2\mathrm{e}^{-4U}\right]\equiv \mathrm{e}^{2V_0}$ in comparison with the star's surface. }\label{surfstrange}
\end{figure}
Finally, we calculated some physical quantities in table \ref{tab:strange}.

\begin{table*}[htb]
\begin{center}\small
\caption{\small Detailed comparison of physical quantities (strange matter equation of state ($\mu = 3p+4B$) and fitted rational potential on the axis). Here, $\tilde p_{\rm c} = p_{\rm c}/B,~\tilde \Omega = \Omega/ B^{1/2},~\tilde M = MB^{1/2},~\tilde R_{\rm circ} =R_{\rm circ}\,B^{1/2}$ and $\tilde J = JB$ are normalized values of the physical quantities. The column labelled AKM shows numerically calculated values, whereas BT[$n$] refers to values of a 2$n$-fold B\"acklund solution. $\Delta$BT[$n$] labels the relative difference between BT[$n$] and AKM.}\label{tab:strange}
\begin{tabular}{cccccccccc}\br
Model  &\hspace{-3.5mm}  AKM    &\hspace{-3.5mm}  BT[1]   &\hspace{-3.5mm}   BT[2]   &\hspace{-3.5mm}   BT[3]   &\hspace{-3.5mm}   BT[4]  &\hspace{-3.5mm}    $\Delta$BT[1]   &\hspace{-3.5mm} $\Delta$BT[2]  &\hspace{-3.5mm}  $\Delta$BT[3]   &\hspace{-3.5mm}  $\Delta$BT[4] \\\mr
$\tilde p_{\rm c} 	$	&\hspace{-3.5mm} 2.00\\
$r_{\rm p}/r_{\rm e}$	&\hspace{-3.5mm} 0.500 \\
$\tilde \Omega 	$	&\hspace{-3.5mm} 3.43 &\hspace{-3.5mm}  4.20 &\hspace{-3.5mm}  3.57 &\hspace{-3.5mm}  3.48 &\hspace{-3.5mm}  3.45 &\hspace{-3.5mm}  2.2e-01 &\hspace{-3.5mm}  4.0e-02 &\hspace{-3.5mm}  1.5e-02 &\hspace{-3.5mm}  4.3e-03\\
$\tilde M 	$		&\hspace{-3.5mm} 0.0355 &\hspace{-3.5mm}  0.0293 &\hspace{-3.5mm}  0.0349 &\hspace{-3.5mm}  0.0354 &\hspace{-3.5mm}  0.0355 &\hspace{-3.5mm}  1.8e-01 &\hspace{-3.5mm}  1.6e-02 &\hspace{-3.5mm}  3.6e-03 &\hspace{-3.5mm}  8.3e-05\\
$\tilde J 	$		&\hspace{-3.5mm} 1.10e-03 &\hspace{-3.5mm}  7.12e-04 &\hspace{-3.5mm}  1.06e-03 &\hspace{-3.5mm}  1.92e-03 &\hspace{-3.5mm}  1.10e-03 &\hspace{-3.5mm}  3.5e-01 &\hspace{-3.5mm}  4.1e-02 &\hspace{-3.5mm}  9.6e-03 &\hspace{-3.5mm}  3.3e-04\\
$\tilde R_{\rm circ}$ 		&\hspace{-3.5mm} 0.141 &\hspace{-3.5mm}  0.129 &\hspace{-3.5mm}  0.139 &\hspace{-3.5mm}  0.140 &\hspace{-3.5mm}  0.141 &\hspace{-3.5mm}  8.4e-02 &\hspace{-3.5mm}  1.9e-02 &\hspace{-3.5mm}  7.8e-03 &\hspace{-3.5mm}  2.4e-03\\
$Z_{\rm p} 	$	&\hspace{-3.5mm} 0.726 &\hspace{-3.5mm}  0.726 &\hspace{-3.5mm}  0.726 &\hspace{-3.5mm}  0.726 &\hspace{-3.5mm}  0.726\\
$Z^{\rm f}_{\rm eq} $	&\hspace{-3.5mm} -0.302 &\hspace{-3.5mm}  -0.411  &\hspace{-3.5mm} -0.326 &\hspace{-3.5mm}  -0.312 &\hspace{-3.5mm}  -0.305 &\hspace{-3.5mm}  3.6e-01 &\hspace{-3.5mm}  7.7e-02 &\hspace{-3.5mm}  3.1e-02 &\hspace{-3.5mm}  8.9e-03\\
$Z^{\rm b}_{\rm eq}$	&\hspace{-3.5mm} 2.28 &\hspace{-3.5mm}  2.17&\hspace{-3.5mm}  2.24 &\hspace{-3.5mm}  2.26 &\hspace{-3.5mm}  2.27 &\hspace{-3.5mm}  5.0e-02 &\hspace{-3.5mm}  1.9e-02 &\hspace{-3.5mm}  9.8e-03 &\hspace{-3.5mm}  3.7e-03\\\br
\end{tabular}
\end{center}
\end{table*}
\normalsize

\end{appendix}

\ack{This work has been supported by the DFG under
GRK1523.}

\References

\harvarditem{{Ansorg} et~al.}{2003}{akm03}
{Ansorg} M, {Kleinw{\"a}chter} A \harvardand\ {Meinel} R  2003 {\em Astron.
  Astrophys.} {\bf 405}~711--21.

\harvarditem{{Belinski} \harvardand\ {Zakharov}}{1978}{bz78}
{Belinski} V~A \harvardand\ {Zakharov} V~E  1978 {\em Zh. Eksp. Teor. Fiz.}
  {\bf 75}~1953--71.

\harvarditem{{Berti} \harvardand\ {Stergioulas}}{2004}{bs04}
{Berti} E \harvardand\ {Stergioulas} N  2004 {\em Mon. Not. R. Astron. Soc.}
  {\bf 350}~1416--30.

\harvarditem{{Ernst}}{1968a}{e68a}
{Ernst} F~J  1968a {\em Phys.\ Rev.} {\bf 167}~1175.

\harvarditem{Ernst}{1968b}{e68b}
Ernst F~J  1968b {\em Phys. Rev.} {\bf 168}~1415--17.

\harvarditem{{Fodor} et~al.}{1989}{fhp89}
{Fodor} G, {Hoenselaers} C \harvardand\ {Perj{\'e}s} Z  1989 {\em \JMP} {\bf
  30}~2252--57.

\harvarditem{{Galassi} et~al.}{2009}{gsl}
{Galassi} M, {Davies} J, {Theiler} J, {Gough} B, {Jungman} G, {Alken} P,
  {Booth} M \harvardand\ {Rossi} F  2009 {\em GNU Scientific Library Reference
  Manual - Third Edition (v1.12)} (Network Theory Ltd).

\harvarditem{{Harrison}}{1978}{h78}
{Harrison} B~K  1978 {\em \PRL} {\bf 41}~1197--1200.

\harvarditem{{Hauser} \harvardand\ {Ernst}}{1981}{he81}
{Hauser} I \harvardand\ {Ernst} F~J  1981 {\em \JMP} {\bf 22},~1051--1063.

\harvarditem{{Kordas}}{1995}{k95}
{Kordas} P  1995 {\em \CQG} {\bf 12}~2037--44.

\harvarditem{{Kramer}}{1980}{k80}
{Kramer} D  1980 in {\em Abstracts of contributed papers} vol 1, Proc. of the 9th Int. Conf. on General Relativity and
  Gravitation, Jena ~42--43.

\harvarditem{{Kramer} \harvardand\ {Neugebauer}}{1968}{kn68}
{Kramer} D \harvardand\ {Neugebauer} G  1968 {\em Commun. Math. Phys.} {\bf
  10}~132--39.

\harvarditem{{Kramer} et~al.}{1991}{knm91}
{Kramer} D, {Neugebauer} G \harvardand\ {Matos} T  1991 {\em \JMP} {\bf
  32}~2727--30.

\harvarditem{{Lichtenstein}}{1933}{l33}
{Lichtenstein} L  1933 {\em Gleichgewichtsfiguren Rotierender Flüssigkeiten}
  (Berlin: Springer).

\harvarditem{{Lindblom}}{1992}{l92}
{Lindblom} L  1992 {\em Phil. Trans. R. Soc. Lond. A} {\bf 340}~353--64.

\harvarditem{{Maison}}{1978}{m78}
{Maison} D  1978 {\em \PRL} {\bf 41}~521--22.

\harvarditem{{Manko} \harvardand\ {Ruiz}}{1998}{mr98}
{Manko} V~S \harvardand\ {Ruiz} E  1998 {\em \CQG} {\bf 15}~2007--16.

\harvarditem{Manko \harvardand\ Sibgatullin}{1993}{ms93}
Manko V~S \harvardand\ Sibgatullin N~R  1993 {\em \CQG} {\bf 10}~1383--1404.

\harvarditem{Maucher}{2008}{m08}
Maucher F  2008 {\em L{\"o}sungen der {E}rnst {G}leichung mit rationalem
  {A}chsenpotential}, Master's thesis, Friedrich-Schiller-Universit{\"a}t, Jena.

\harvarditem{{Meinel} et~al.}{2008}{maknp08}
{Meinel} R, {Ansorg} M, {Kleinw{\"a}chter} A, {Neugebauer} G \harvardand\
  {Petroff} D  2008 {\em {Relativistic Figures of Equilibrium}} (Cambridge: Cambridge University Press).

\harvarditem{{Meinel} \harvardand\ {Neugebauer}}{1995}{mn95}
{Meinel} R \harvardand\ {Neugebauer} G  1995 {\em \CQG} {\bf 12}~2045--50.

\harvarditem{Nelder \harvardand\ Mead}{1965}{nm65}
Nelder J~A \harvardand\ Mead R  1965 {\em Comput. J.} {\bf
  7}~308--13.

\harvarditem{{Neugebauer}}{1979}{n79}
{Neugebauer} G  1979 {\em \JPA} {\bf 12}~L67--L70.

\harvarditem{{Neugebauer}}{1980{\em a}}{n80a}
{Neugebauer} G  1980{\em a} {\em \JPA} {\bf 13}~L19.

\harvarditem{{Neugebauer}}{1980{\em b}}{n80b}
{Neugebauer} G  1980{\em b} {\em \JPA} {\bf 13}~1737--40.

\harvarditem{Neugebauer}{1996}{n96}
Neugebauer G  1996 in {\em General
  Relativity} edited by  Hall G~S \harvardand\ Pulham J~R,  SUSSP Publications and Institute of Physics Publishing, Edinburgh University and The Institute of Physics, London ~61--81.

\harvarditem{{Neugebauer} \harvardand\ {Hennig}}{2009}{nh09}
{Neugebauer} G \harvardand\ {Hennig} J  2009 {\em Gen. Rel. Grav.} {\bf
  41}~2113--30.

\harvarditem{{Neugebauer} \harvardand\ {Kramer}}{1983}{nk83}
{Neugebauer} G \harvardand\ {Kramer} D  1983 {\em \JPA} {\bf 16}~1927--36.

\harvarditem{{Neugebauer} \harvardand\ {Meinel}}{2003}{nm03}
{Neugebauer} G \harvardand\ {Meinel} R  2003 {\em \JMP} {\bf 44}~3407--29.

\harvarditem{{Nozawa} et~al.}{1998}{nozawa}
{Nozawa} T, {Stergioulas} N, {Gourgoulhon} E \harvardand\ {Eriguchi} Y  1998
  {\em Astron. Astrophys.} {\bf 132}~431--54.

\harvarditem{{Pappas}}{2009}{p09}
{Pappas} G  2009 {\em J. Phys.: Conf. Ser.} {\bf 189}~012028.

\harvarditem{{Sibgatullin}}{1991}{s91}
{Sibgatullin} N~R  1991 {\em {Oscillations and Waves in Strong Gravitational
  and Elec\-tromagnetic Fields}} (Berlin: Springer).

\harvarditem{{Stute} \harvardand\ {Camenzind}}{2002}{sc02}
{Stute} M \harvardand\ {Camenzind} M  2002 {\em Mon. Not. R. Astron. Soc.} {\bf
  336}~831--40.

\endrefs

\end{document}